\documentclass[aps,pre,twocolumn,nofootinbib]{revtex4}
\usepackage{graphicx}
\usepackage[utf8]{inputenc}
\usepackage{color}
\newcommand\strongdisorder{{ $R_{exc}/R_i=0.63$ }}
\newcommand\lowdisorder{{ $R_{exc}/R_i=1.5$ }}
\newcommand{\change}[1]{ #1 }
\begin{document}
\title{Phase field study of the effective fracture energy increase during dynamic crack propagation in disordered heterogeneous materials.}
\author{Hervé Henry}
\affiliation{  Laboratoire Physique de la Matière Condensée (PMC), CNRS,  École Polytechnique-Institut Polytechnique de  Paris, 91120 Palaiseau, France}

\begin{abstract}

{The propagation of a 3D crack in an heterogeneous material is studied using a phase field model. It is shown that in the case of randomly distributed inclusions of soft material in a matrix, the nature of the distribution has little effect on the effective elastic properties. On the opposite it affects significantly crack propagation. The less uniform distribution leads to higher thresholds for crack propagation.}
\end{abstract}
\maketitle
\section{Introduction}

Despite the remaining challenges \cite{Kammer2024,Wang2023,Livne2007} in understanding  brittle and  ductile crack propagation in homogeneous materials such as glass or polymers, the theoretical tools that have been developed over the last decades allow to  predict well the evolution of a crack in an homogeneous material at both low and high speeds. On the contrary in the case of heterogeneous materials this is not case. Indeed, the LEFM theory relies on the  scale separation between the macroscopic material size (that can be of the order of mm to km) and the microscopic process zone at the crack tip where  the irreversible material breaking process takes place. When considering an heterogeneous  elastic material, the description of its elastic properties is given by using a so called \textit{representative elementary volume} that is such that the elastic properties of this volume are the properties of the material at large scale. This approach relies on the scale separation between the microstructure and the macroscopic scale at which the elasticity problem is solved.  

Such a scale separation is not present in the case of cracks: the material breaking takes place in the \textit{process zone} that is often nanometric and much smaller than the characteristic size of the microstructure. As a result an heterogeneity of material properties (for instance fracture toughness) will affect the crack propagation significantly. For instance a high toughness region may prevent the crack from advancing locally and may pin the crack front in 3D or simply stop the crack in 2D.
This scenario has already been discussed in the case of fracture toughness heterogeneities and it has been shown, in the limit of small deformation of the front that heterogeneities can lead to higher fracture energies\cite{Ortiz1994,Lebihain,Lebihain2023,Lebihain2021}. In addition to these work where a randomly distributed population of inclusions were considered, other work have focused on architected structures such as stripes and other recent work have focused on the effects of fracture toughness anisotropy.\cite{Brach2019}
  
\begin{figure}
   \includegraphics[width=0.45\textwidth]{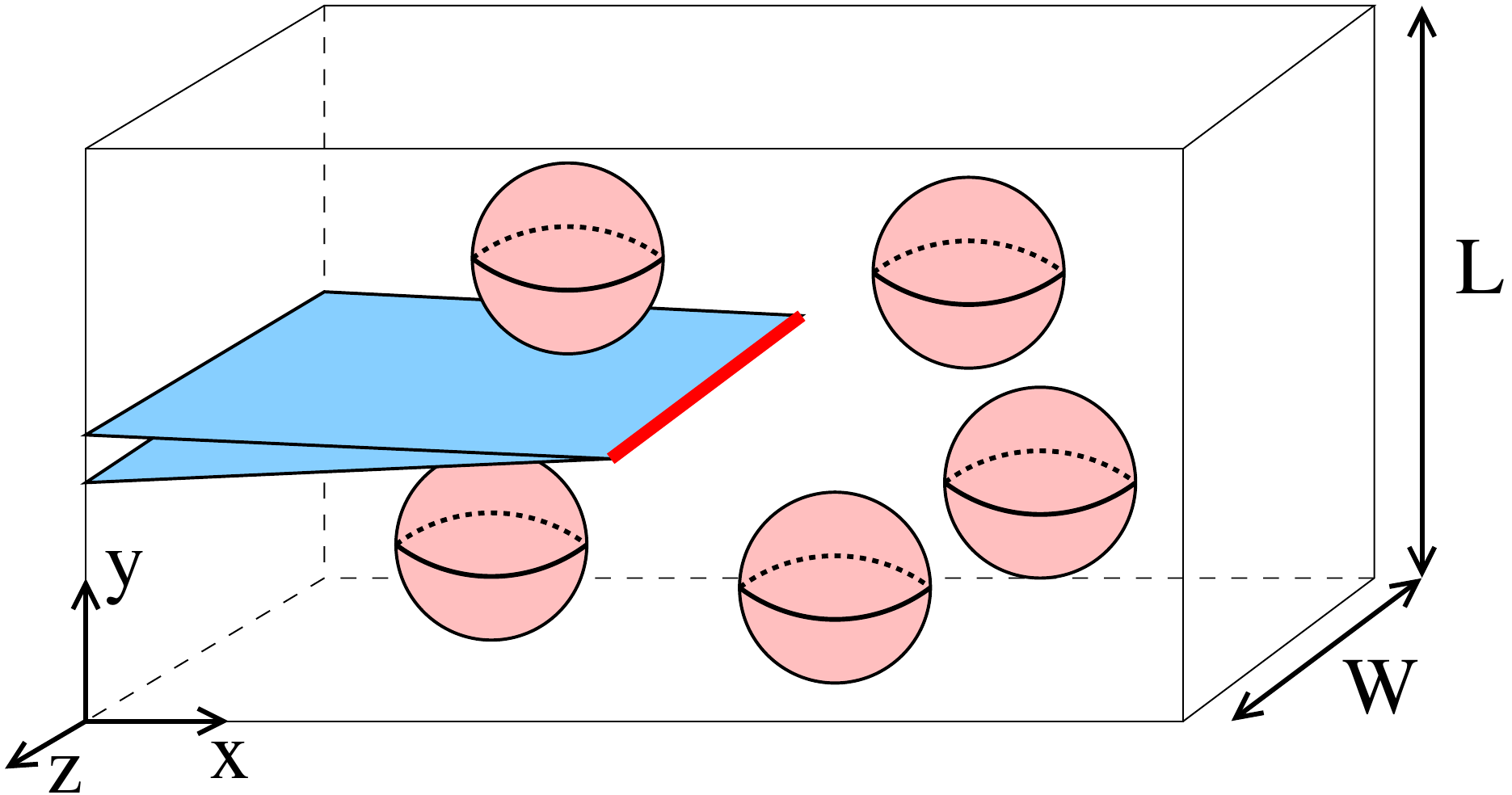}
   \caption{Schematic view of the system considered. A crack is propagating along the x axis in a material under load (imposed displacement at $y=\pm L/2$) in a material that is composed of a matrix and spherical inclusions with  lower elastic constants.  Boundary conditions at $z=\pm W/2$ are periodic. A treadmill geometry is used to follow the crack propagation along the x axis. \change{In most simulations described here $L=120$, $W=120$ and the radius of inclusions is $R_i=12$}\label{fig:setup}}
 \end{figure}

The effect of elastic moduli heterogeneities  has been less studied and the works have been limited to single inclusions or to  architected geometries such as stripes.  In the case of single inclusion it has been shown that both hard and soft inclusions  can pin a crack front in 2D and 3D if  the geometry is favorable\cite{Henry2024,Clayton2014}. However in an actual material there are many inclusions present that affect both the material properties and the crack propagation. Here, in the spirit of building a bridge between structure and properties. The propagation of a crack front in an heterogeneous material composed of a matrix and of spherical inclusions with a lower elastic moduli is studied.  The link between the spatial organization of inclusions and its  apparent fracture energy and effective elastic moduli is discussed. To this purpose phase field simulations of crack propagation in an infinite strip geometry (see fig. \ref{fig:setup})  have been performed with varying distribution of soft inclusions (randomly distributed, randomly distributed with a minimal center to center distribution that leads to an almost hyper-uniform distribution). The results indicate that the less uniform distribution of inclusions (for which they can form clusters)  leads to an apparent higher fracture toughness, especially at the onset of crack propagation while it does not affect effective material properties\cite{Heitkam2016}.

 The paper is organised as follows. First the way spherical inclusions can be arranged in the material is discussed and the way effective elastic moduli are computed is discussed. It is seen that in the case of a geometry where inclusions are in a matrix, the elastic moduli depend on the sole volume fraction of the inclusions. Second the phase field model for crack propagation is briefly described. Third numerical results are presented and discussed. Finally concluding remarks are given. 

\section{Elastic properties of randomly arranged  inclusions}
\begin{figure}
  \centerline{\includegraphics[width=0.45\textwidth]{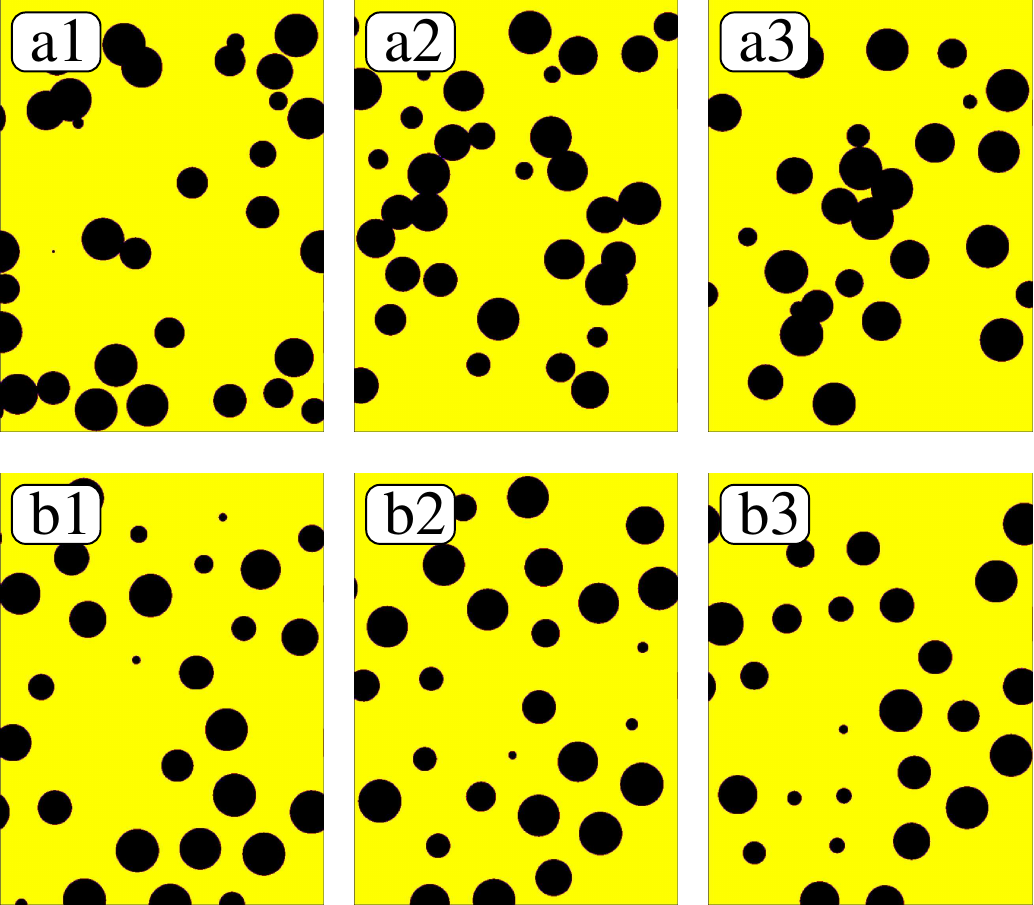}}
  \caption{\label{fig:inclusions}Top row (a1, a2, a3) cuts of the domain with the inclusion for \strongdisorder at 3 different $y$ equally separated by 2.5 inclusion radius . Bottom row (b1, b2, b3) cuts of the domain with the inclusion for \lowdisorder  at 3 different $y$ equally separated by 2.5 inclusion radius.  }
\end{figure}

   Considering a matrix material, where a given volume fraction  is occupied by inclusions, there are many ways the spacial organisation of the material can be varied. The shape of inclusions and where they are positioned can be changed. For instance, one can consider spherical inclusions positioned on a lattice similarly to what is observed in a crystal. Ellipsoidal  inclusions randomly  positioned in regularly spaced layers as in liquid crystals is another possibility. One can also consider spherical inclusions  randomly placed in order to satisfy a given set of criteria. The number  of different possibilities is large even if only spatial organisations inspired by small scale structure of materials are considered. The aim of this work is to show that the nature of the spatial arrangement of inclusions can affect fracture properties while affecting much less significantly its elastic properties. 

   For the sake of simplicity random arrangement of inclusions are considered. They  have the advantage of being isotropic and homogeneous at large scale. In addition their properties have already been discussed in the framework of homogenization theory\cite{Christensen1993}. In the case of randomly distributed inclusions, one way to \textit{tune}  the disorder is to vary the minimum distance between centers. 
   Indeed,  if this minimum center to center distance  is small, clusters of inclusions next to each other will appear, sometimes containing  a large number of inclusions. On the opposite if one imposes a large center to center minimum distance, inclusions will organize with a constant number density and very small fluctuations of number density of inclusions will be observed. Such arrangement can be characterized by the number density of grains $n$ and by the minimal distance between grain centers $2R_{exc}$. From these two parameters  a \textit{virtual} volume fraction $\Phi_{exc}$ can be computed. It corresponds to the volume fraction of grains with the same number density and a radius of $R_{exc}$. 
   \begin{equation}
     \Phi_{exc}=n\times \frac{4}{3}\pi R_{exc}^3
   \end{equation}
   The maximal values of $\Phi_{exc}$ in the case of random arrangement of inclusions is $\approx 0.64$ \cite{grains1} and there exists efficient algorithms that allow to build populations of grains with such properties\cite{Zinchenko1994,Torquato2000}. Here we use an  algorithm where the radius of hard spheres is slowly increased during the course of a Monte Carlo simulation\footnote{Particle centers are randomly  moved one after the other. The motion is rejected if it leads to an overlap. Periodically the radius of spheres $R_{exc}$ is increased to reach the maximal value the current particle distribution is  compatible with }. It allows to reach $\Phi_{exc}\approx 0.64$. In the course of this work the actual  volume fraction occupied by the spheres \change{of fixed radius $R_i$ has been chosen to be $\Phi=0.2$, neglecting possible overlap of the spheres. Their radius  $R_i$ was taken equal to 12} and the minimal distance between them was varied between 15.14 ($<2R$) (value for which there is a small amount of  overlap) and 35.18  that corresponds to $\Phi_{exc}\approx0.63$. Typical distributions are presented in fig.\ref{fig:inclusions} for these two value using  cuts at different values of $y$. One can clearly see the difference between these two distributions~: for \strongdisorder there are clusters of inclusions whose size is of the same order as the system size. On the opposite for \lowdisorder such clusters are not present. It must be also noted that the amount of overlap is small for \strongdisorder. 

   In order to estimate the effect of these distributions of inclusions on the elastic properties of the material we consider a piece of material at equilibrium  under fixed displacement at the top and bottom boundary and compute the elastic energy stored  in slabs that spans the whole system in the $y$ and $z$ direction and has a length of the radius of an inclusion in the $x$ direction.    The elastic moduli of inclusions are taken 10 times smaller than the elastic moduli of the matrix. The ratio of this value  with the elastic energy stored in an homogeneous material with the same properties as the matrix  and with the same imposed displacement is computed for the two kinds of disorder presented  here. \change{This elastic energy ratio } has an average value of $\approx0.66$ in both cases. And the relative standard deviation  of the values is about 0.06 for \strongdisorder and 0.05 for \lowdisorder. Hence, the effect of the nature of disorder with the chosen distributions is small. When comparing these simple estimations with literature, we find that  the value of the ratio is, as expected,  slightly larger  than  the one predicted theoretically for voids\cite{Christensen1993} which is $\approx0.6$. The fact that the spatial organisation of voids does not affect the effective elastic moduli is in agreement with \cite{Heitkam2016} \change{where the Young's modulus of a matrix with different periodic arrangements of voids was found to be independent of the nature of the voids. In \cite{Heitkam2016} it was also found that (with the exception of the simple cubic lattice) that the effective Poisson ratio was not changed significantly with the nature of the organisation of voids}. 

\section{Phase field model}

The phase field model used here was originally introduced in \cite{KKL} \change{for mode III crack, it was later extended to general loading in \cite{henry-04}}. It  has been shown to reproduce well LEFM results in the case of quasi static crack propagation\cite{Corson2009} and in the case of dynamic fracture propagation\cite{henry2008}. It relies on the introduction of a phase variable $\phi$ that describes the state of the material intact($\phi=1$) or broken ($\phi=0$).  The total energy of the system writes then:
\begin{equation}
  \mathcal{F}=\int \sqrt{\scriptstyle \Gamma/w_\phi}(V(\phi)-\epsilon_c^2 g(\phi))+\sqrt{\scriptstyle \Gamma w_\phi}(\nabla \phi)^2+g(\phi)\mathcal{E}_{el} dV 	
\end{equation}
where and $g(\phi)=4\phi^3-3\phi^4$ couples the phase field with the elastic field. With  $V(\phi)=\phi^2(1-\phi)^2$,  
$(V(\phi)-\epsilon_c^2 g(\phi))$ is a tilted double well potential with minima at $\phi=0$ and 1. $\mathcal{E}_{el}$ is the elastic energy density. \change{Here the model parameters $\Gamma$, $\epsilon_c$ and $w_\phi$ are taken equal to 1 (\textbf{check}). With this choice of parameters, the total fracture energy, twice the surface energy, is... Multiplying $\gamma$ by a factor of 2 will increase the fracture energy by 2.   Multiplying $w_\phi$by a factor of 2 will double the phase field interface thickness. } 
It must be noted that with this model there is no strain-softening since the coupling function keeps the minima   at  $\phi=0$ and 1. Moreover, with such model nucleation of cracks is not possible unless unrealistically high loads are applied. This is in contrast with models such as the one discussed in \cite{Maurini}.  The elastic energy density writes:
\begin{equation}
  \mathcal{E}_{el}=\frac{\lambda(\mathbf{x})}{2}(\mathrm{tr}\varepsilon)^2+\mu(\mathbf{x})\mathrm{tr}(\varepsilon^2)
\end{equation}
where $\varepsilon$ is the strain tensor and $\lambda$ and  $\mu$ are space dependant Lam\'e Coefficients.  As in \cite{Henry2024}, $\lambda$ and  $\mu$ are taken to be equal to 0.1 in inclusions while they are taken equal to 1 in the matrix.\change{ It must be noted that  $\Gamma$, $\epsilon_c$ and $w_\phi$ are unchanged in the matrix and the inclusions}.

\change{The motion  of the crack  is described by an evolution equation, in the same spirit as in \cite{miehe}: a relaxation equation. The   evolution of  elastic fields corresponds to the momentum balance taking into account the presence of the crack and result in the normal stress wave equation with both inclusions and matrix sharing the same mass density. Using the functional derivative $\delta· /\delta \phi$ this translates into the following evolution equation:}
\begin{eqnarray}
  \partial_t \phi&=& -\beta \frac{\delta \mathcal{F}}{\delta \phi} \times f({\delta \mathcal{F}}/{\delta \phi},\varepsilon)\\
  \partial_{tt} u_i&=&- \frac{\delta \mathcal{F}}{\delta u_i}
\end{eqnarray}
where $u_i,\ i\in\{x,\ y,\ z\}$ are the displacement fields and $\beta$ is a positive real that is proportional to the energy dissipation at low speeds\cite{henry2008}.  The positive function $f$ has the following form:
\begin{equation}
  f=\left\{
  \begin{array}{l}
    \mathrm{max}(0,\frac{\displaystyle \delta F/\delta \phi  -g'(\phi)K_{L}( tr \varepsilon)^2  }{\displaystyle  \delta F/\delta \phi}) \mbox{ if }  tr \varepsilon <0\\
  0\mbox{ if } {\delta \mathcal{F}}/\delta \phi < 0\\
  1 \mbox{ otherwise }
  \end{array}
  \right.
\end{equation}
It  guarantees  that the elastic energy due to  compression  will not contribute to crack propagation and  that there will be no crack healing. 

 The model is simulated using finite differences in space and a Verlet time stepping scheme that guarantees that the discretized total free energy  is actually decreasing numerically.  The grid spacing used is $\Delta x=0.3$ and it has been  checked  that decreasing it by a factor of 2 (i.e. doubling the resolution) does not affect the results significantly.  Same checks were performed for the timestep and for the fracture interface thickness $w_\phi$. 

 All the simulations discussed here are performed with the same protocol~: the elastic configuration of  a pre cracked\footnote{\change{The phase field is taken equal to $\phi(x,y)=1.- \exp(-(y-L/2)^2/9.)\times(1+tanh((L-x)/1.5))/2$, where $L=120$ is the system extension along the $y$ axis and $240$ is the length of the simulation box along the $x$ axis.}}  homogeneous  elastic material   is  computed under a given strain ($\Delta_y=1$),\change{ that corresponds to pure mode I loading}. The elastic field are rescaled to reach the prescribed load.  Then the simulation of the model starts with elastic heterogeneities. The motion of the crack is followed by computing the most advanced point in each $xy$ plane of the grid locally (local maxima of $x$ of the crack interface defined as the isosurface $\phi=0.5$) using a simple linear interpolation scheme. A treadmill is used to follow the crack propagation  over long distances. It has been  checked that doubling the domain size along the x axis does not affect the results, indicating that the treadmill is \textit{free} of significant finite size bias.

 \change{The prescribed  displacement at $y=\pm L/2$: $\pm \Delta_y/2$  is the parameter  used to tune the elastic energy available for crack propagation: for each value of $\Delta_y$ a simulation is run, keeping this parameter fixed,  over a time that corresponds to the propagation of the crack over a distance of a few $L$. The velocity of the crack is the average velocity of the crack front after a short transient that corresponds to the advance of the crack over a distance of$\approx L$. The zero velocity fracture energy is defined as the highest  elastic  energy per unit length  in the intact system for which no steady crack propagation is observed. }

\begin{figure}
  \centerline{\includegraphics[width=0.45\textwidth]{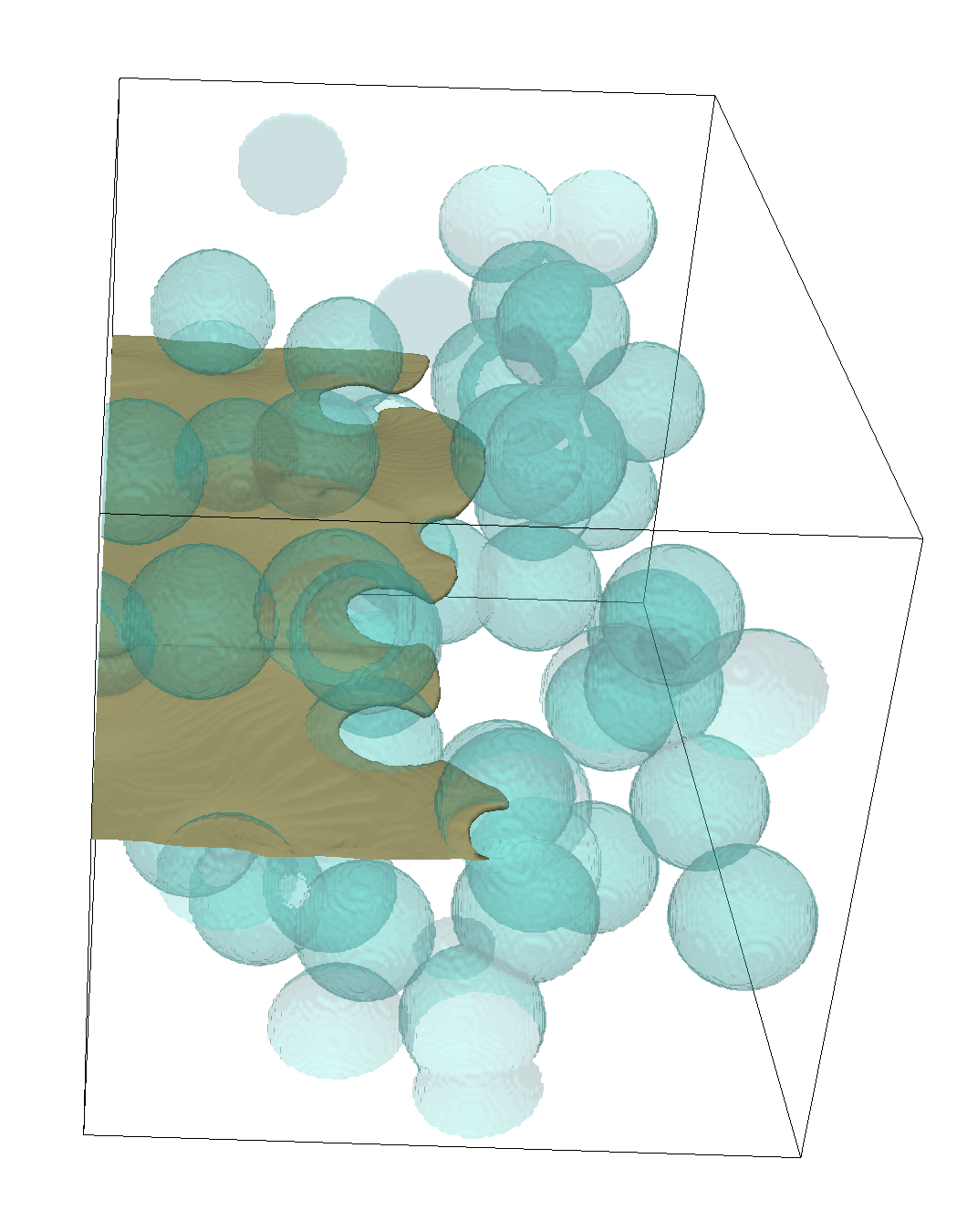}}
  \caption{typical perspective view of the crack front together with inclusions \change{in the frame of the unstrained material}. Only a part of the simulation domain along the $x$ axis is shown. The crack is propagating from left to right.\label{fig3d}. The deviation of the front from the midplane is much smaller than the size of the inclusions and its surface remains approximately flat. }
\end{figure}

\section{Results }
\begin{figure}
 \centerline{\includegraphics[width=0.45\textwidth]{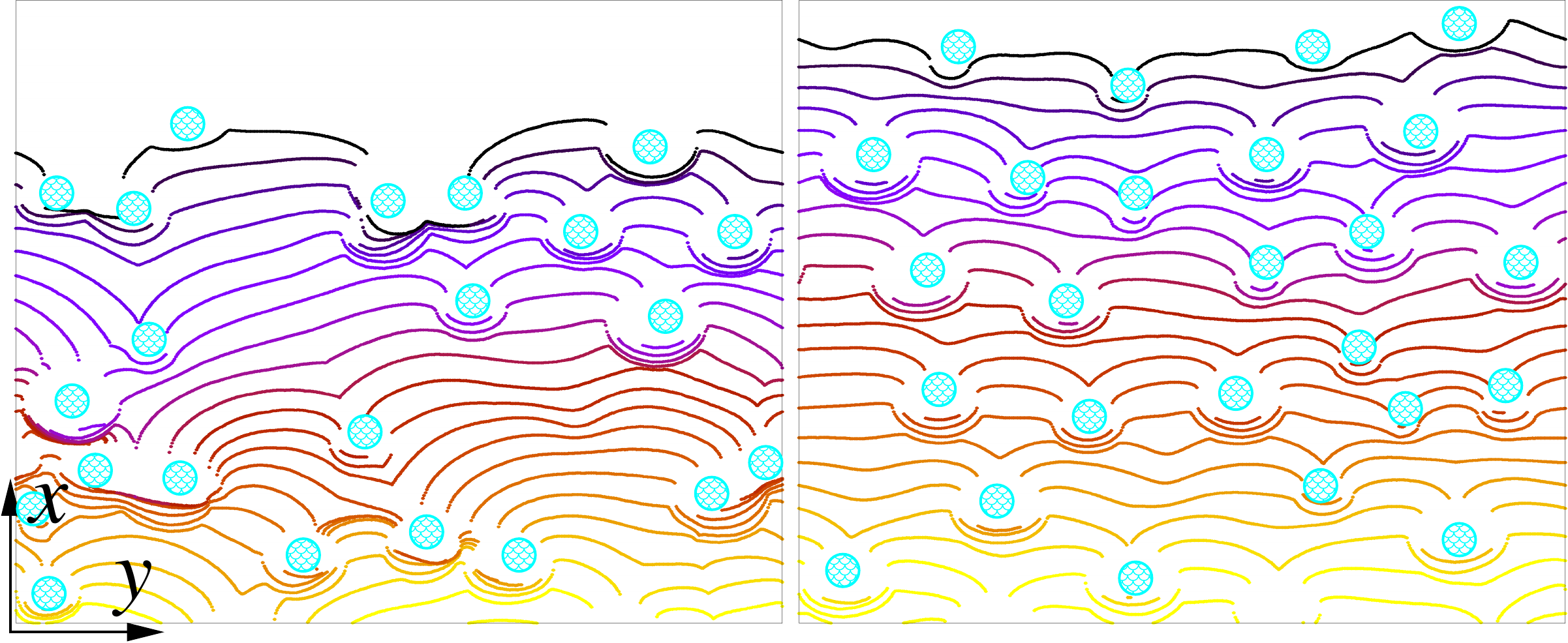}}
 \caption{\label{figlapse} Top view of crack front \change{at different evenly spaced times}  for both disorders considered here: \strongdisorder(left) and \lowdisorder (right). \change{ The color-scale (gray-scale) correspond to the time at which the crack front is observed. It goes from yellow (light gray using a gray scale)  to black as time increases. The closely spaced lines correspond to local slow down of the crack front.} While the number of inclusions encountered by each front are very close, one can see that there are more clusters of inclusions in one case and that the front is much more distorted.}
\end{figure}

The elastic moduli  heterogeneities affect the crack propagation in many aspects. It was shown in \cite{Henry2024,Clayton2014} that a single heterogeneity could trap a crack if properly positioned with respect to the plane of propagation of a crack. In the context of materials,it is not known   where a crack will propagate. Therefore if the sought after effect is to trap a crack with heterogeneities, heterogeneities must be randomly distributed in the material. Their presence will obviously affect the effective  elastic  properties of the material. They will also affect crack propagation. This was discussed in the case of toughness heterogeneities in\cite{Ortiz1994,Lebihain} for instance.  Similar behaviour in the case of elastic moduli heterogeneities is  seen in fig.\ref{fig3d} where a perspective view of a crack propagating in a random material is shown for both kind of disorders together with the inclusions. One can see that the front is deformed due to the presence of inclusions and one can see that the front deformation is locally similar to what was described in  \cite{Henry2024,Clayton2014} where a single inclusion was considered. With the parameters chosen here, there are more than one occurrence of front-inclusion interaction that can be seen at a given time. In the following we discuss qualitatively the possible cooperative effects that can occur and quantify the effect of a population of inclusions.

\begin{figure}
  \centerline{\includegraphics[width=0.45\textwidth]{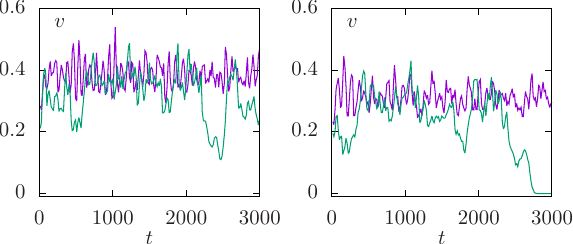}}
  \caption{\label{figtracesvitesses} Plots of the \change{ instantaneous crack front velocity along the $x$ axis averraged over the simulation domain  thickness} as a function of time for two applied load:\change{ $G/G_c\approx1.7$ on the left panel and $G/G_c\approx1.4$ on the right panel. On each panel the purple line corresponds to a \lowdisorder and the green one to a \strongdisorder}.  One can see that for \lowdisorder  there are \textit{moderate} irregularities in the crack average velocity. When \strongdisorder  the oscillations are of much larger amplitude and in the case of the low load (right panel) they are such that the crack eventually stops after propagating over a long distance ( $\approx 80$   inclusion radius).}
\end{figure}

In fig \ref{figlapse}, a sequence of top  views of the crack front together with the approximate position of  inclusions is shown for the two kind of disorder considered here. The front of the crack is no longer straight and presents a complex structure that is reminiscent  to what can be seen in the context of propagating fronts in disordered media (avalanches, moving contact line and fractures)\cite{Rosso,santucci,Bares,Dubois,Wiese}. The similarity is clearer in the case of \textit{strong} disorder: i.e. when  \strongdisorder. Indeed in this case the front distortion is much more visible while in the case of \textit{small} disorder, the distortion of the front are weaker (of the same order of magnitude as the size of inclusions). This illustrates the fact that in the case \strongdisorder, inclusions can have cooperative effects on the propagation of crack. This translates into the fact that when such cooperative effects the pinning of the crack front at a group of inclusions is longer. As a result one expects that the average crack front velocity will present an intermittent behaviour. 

 This is illustrated in fig. \ref{figtracesvitesses} where, the average crack front velocity is plotted as a function of time for two loads (the lower on the right panel). For both values of load the crack velocity variations are larger  for \strongdisorder. For the lowest load and strongest disorder, the crack can be arrested by the inclusions after propagating over a long time at relatively high velocities ( $\approx 0.24 c_s$). For \lowdisorder, the variations around the average velocity are much lower and the motion of the crack front is smoother. Finally, one can see that for both values of the load, the average crack front velocity is smaller for \strongdisorder while the elastic moduli of the effective material are equal. This shows  that the clusters of inclusions that are present for    \strongdisorder   have dramatic effects on the propagation of cracks. Hence the nature of disorder has a significant effect on the fracture propagation threshold. 

\begin{figure}
  \includegraphics[width=0.5\textwidth]{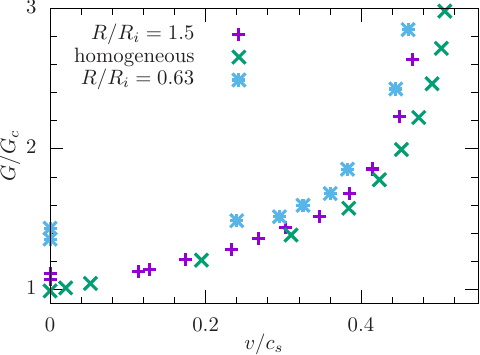}
  \caption{Fracture energy as a function of the crack front speed averaged \change{over time and space} for different \change{kind of disorder and load}. The crack energy is computed using the formula of\cite{Freund} with an effective Young's modulus computed numerically for a Representative Elementary Volume (REV) for each case (homogeneous and distributions. One can see that for $R/R_i=0.63$ there is a \textit{jump} in crack velocity. We were not able to observe cracks propagating with an average speed of less than $\approx 0.24 c_s$. \label{figenergies} }
\end{figure}

The effect of the disorder on the \textit{effective} fracture energies is shown in fig.\ref{figenergies}. The fracture energy is plotted as a function of crack  velocity for different elastic moduli disorder. To this purpose the fracture energy is computed  \change{ using the  effective elastic moduli $E_{eff}$ that where computed earlier, with the expression  from \cite{Freund}:
\begin{equation}
	G=\frac{\Delta_y^2}{W}E_{eff}\frac{1}{A_I(v)}\label{eq_ERR}
\end{equation}
with $A_I(v)$ a universal function whose expression is given in appendix \ref{App_AI}.}
The fracture energy is higher for the heterogeneous material than for the \textit{equivalent} homogeneous material. In addition, while for the homogeneous material the crack velocity can be as low as a fraction of $c_s$ in the infinite strip geometry, in the case of \strongdisorder there is a minimal velocity close to $0.29 c_s$  below which the crack cannot propagate. This corresponds to an apparent fracture energy $\approx 1.5$ times larger than the fracture energy of the homogeneous material.\change{ In the case of \lowdisorder,  there is a small jump since no crack speed below $\approx 0.1 c_s$ is observed} and the fracture energy at zero velocity is $\approx 1.25$ times larger than the fracture energy of the homogeneous material. As soon as the crack starts to propagate in a \textit{steady state} (that is without stopping)  the difference between the two kind of disorder is less marked and the apparent fracture energy increase when compared to the homogeneous material  is of the order of 10\%. Surprisingly at high crack velocity a significant increase in the apparent fracture energy is seen and it can reach $\approx 40\%$. To better understand this phenomenon we study the fracture surface in this case.

\begin{figure}
  \includegraphics[width=0.5\textwidth]{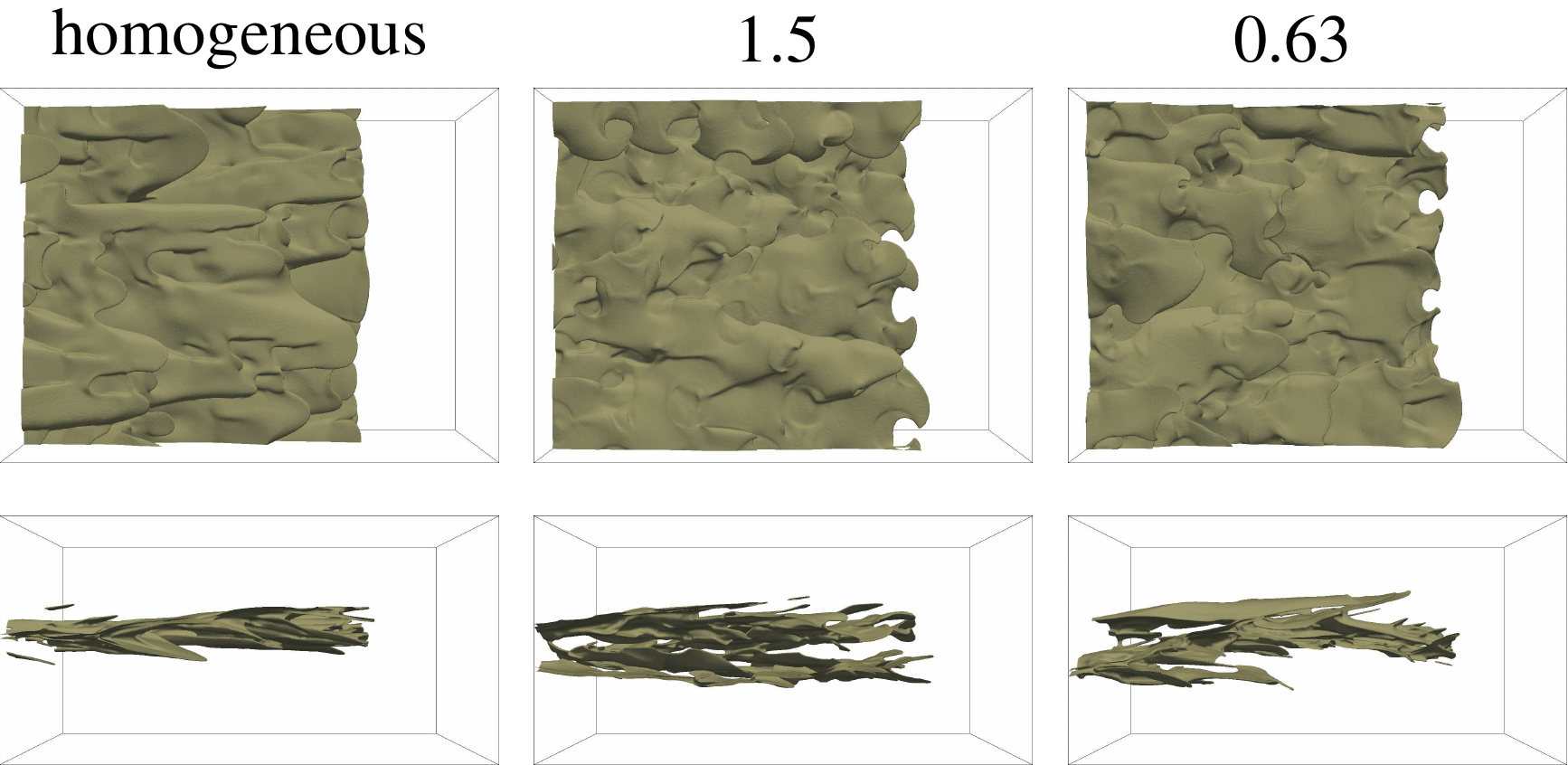}
  \caption{\label{branching} Top and side view of a crack surface propagating at a speed well above the branching threshold. The applied load is such that the total elastic energy stored in the material is the same in all three cases. It corresponds to $G\approx 4.5 G_c$.}
\end{figure}
Indeed at high speed, the crack can present tip splitting and microbranching as already seen in various phase field simulations\cite{henry2008,karmabranch,bleyer}, experimentally\cite{sharon1994} and studied theoretically\cite{addabranching,katzav2007}. This microbranching instability is the origin of the  high apparent fracture energy\cite{Sharon1995} at high crack velocity.  
 In figure \ref{branching}  the fracture surface is represented from above for homogeneous materials, \lowdisorder and \strongdisorder for a given amount of available elastic energy such that in all cases there is a complex branching pattern. These plots illustrate the fact that while in the case of an homogeneous system, the pattern seen from above is visibly oriented  along the crack propagation direction, this is less pronounced in the presence of disorder. Moreover, in the presence of inclusions the crack forms well marked steps (reminiscent of echelon cracks) that can be seen on the side view where it is clear that for disordered systems there are cracks propagating along distinct planes approximately parallel to the $xy$ plane. However these qualitative results have been obtained on too small systems to allow a quantitative analysis as in experiments \cite{Ponsonstat,Dalmasstat}. Comparing the total fracture surface created is possible and indicates, contrarily to what would be expected, that $\approx 15\%$ more fracture surface is created for homogeneous material and \lowdisorder than for \strongdisorder. As a result, the observed increase in fracture energy cannot be attributed to the fact that more fracture surface is created. It is more likely that it is related to the same mechanism as the fracture energy increase observed at lower speeds.

\section{Conclusion and discussion}

Numerical simulations of cracks propagating in elastic  heterogeneous materials have been performed. While the material considered has always been a matrix containing soft inclusions, different kind of disorder have been considered by varying the minimal distance between inclusions keeping their volume fraction constant.   The nature of disorder significantly affects the crack front shape, the average crack speed trace. It has also been found that the fracture effective energy, defined as the fracture needed to have the crack advance without stopping is larger in disordered material and that in the case of strong disorder there can be a discontinuity in crack velocity when increasing the applied  load. Below the threshold the crack is not propagating while  above it propagates at a velocity of $\approx 0.24c_s$. 

These results have been obtained in simulations with large systems. However, the system are such that some of the effects described here are \textit{finite size} effects.

In the case of \lowdisorder the increase in effective fracture energy is apparently not a finite size effect and should be observed in larger systems. It indicates that such materials  have  higher fracture energy than homogeneous materials

In   the case   \strongdisorder,  the existence of large clusters of inclusions  is  leading  to crack arrest. However, infinite  clusters in an infinite medium do not exist below the percolation threshold. Therefore the effect described here should not hold in the large system size limit. However, this behaviour can be observed for much larger systems. Indeed  the results are still valid for larger systems as as been shown by  numerical experiments that were performed with twice larger systems \change{(20 times the  inclusion radius $R_i$ while simulations described here are for 10 times $R_i$)}. And they should hold for larger systems. \change{This is the case, despite the difference of boundary conditions, for  thin plates whose thickness can be only a not too large multiple of $R_i$}.   These results are valid for systems that can be seen as homogeneous from the point of view of elasticity (at large scale). They  indicate that the probability distribution function of cluster sizes is likely to be a key tuning parameter to design materials with increased fracture energy.  

\change{Finally it is worth mentioning that the toughening effect observed here is reminiscent of similar results presented in\cite{Ortiz1994} where the presence of difficult to break inclusions lead to the formation of a cohesive zone and additional energy dissipation at the crack tip. However, in the framework considered here, especially for \strongdisorder, it is difficult to extract an unique characteristic length: it may be the particle size or a characteristic cluster size (if it exists). This indicates that further analysis in this direction is needed.}

\change{
\section*{Data availability}
The supporting data for this article are openly available from the https://entrepot.recherche.data.gouv.fr/ (exact DOI to be added with proofs).

\begin{acknowledgments}
  This work has been supported by the CIEDS contract FRACADDI 
\end{acknowledgments}

}

\appendix
\change{
\section{Expression of $A_I(v)$\label{App_AI}}
In this appendix, the expression of $A_I(v)$ used in eq.\ref{eq_ERR} is given. First the shear wave speed and dilational wave speed  are $c_s=\sqrt{\mu/\rho}$ and $c_d=\sqrt{(\lambda+3\mu)/\rho}$. The following auxiliary functions are defined:
\begin{equation}
	\alpha_d(v)=\sqrt{1-v^2/c_d^2}\mbox{ and } \alpha_s=\sqrt{1-v^2/c_s^2}
\end{equation}
Then the Raileigh wave speed $c_R$ is solution of:
\begin{equation}
	0=D(v)=4\alpha_d\alpha_s-(1+\alpha_s^2)^2
\end{equation}
and $A_I(v)$ writes:
\begin{equation}
	\frac{v^2/c_s^2 \alpha_d(v)}{(1-\nu)D(v)}
\end{equation}
It must be noted that $A_I(v)$ diverges at the Rayleigh wave speed because $D(c_R)=0$ and that $A_I(0)=1$. This implies that the crack cannot reach a velocity higher than $c_R$.
}

\end{document}